\documentclass[conference]{IEEEtran} 
\IEEEoverridecommandlockouts
\usepackage{cite}
\usepackage{amsmath,amssymb,amsfonts}
\usepackage{algorithmic}
\usepackage{graphicx}
\usepackage{textcomp}
\usepackage{xcolor}
\def\BibTeX{{\rm B\kern-.05em{\sc i\kern-.025em b}\kern-.08em
    T\kern-.1667em\lower.7ex\hbox{E}\kern-.125emX}}
\begin{document}

\title{GraphBLAS on the Edge: \\ Anonymized High Performance Streaming of Network Traffic
\thanks{This material is based upon work supported by the Under Secretary of Defense for Research and Engineering under Air Force Contract No. FA8702-15-D-0001, National Science Foundation CCF-1533644, and United States Air Force Research Laboratory and Artificial Intelligence Accelerator Cooperative Agreement Number FA8750-19-2-1000. Any opinions, findings, conclusions or recommendations expressed in this material are those of the author(s) and do not necessarily reflect the views of the Under Secretary of Defense for Research and Engineering, the National Science Foundation, or the United States Air Force. The U.S. Government is authorized to reproduce and distribute reprints for Government purposes notwithstanding any copyright notation herein.}
}

\author{\IEEEauthorblockN{Michael Jones$^1$, Jeremy Kepner$^1$, Daniel Andersen$^2$, Ayd{\i}n Bulu{\c{c}}$^3$, Chansup Byun$^1$,   K Claffy$^2$, Timothy Davis$^4$,  \\ William Arcand$^1$, Jonathan Bernays$^1$, David Bestor$^1$, William Bergeron$^1$, Vijay Gadepally$^1$,  Micheal Houle$^1$, \\Matthew Hubbell$^1$,  Hayden Jananthan$^1$, Anna Klein$^1$, Chad Meiners$^1$, Lauren Milechin$^1$, Julie Mullen$^1$, \\ Sandeep Pisharody$^1$, Andrew Prout$^1$,  Albert Reuther$^1$, Antonio Rosa$^1$, Siddharth Samsi$^1$, \\ Jon Sreekanth$^5$, Doug Stetson$^1$, Charles Yee$^1$, Peter Michaleas$^1$
\\
\IEEEauthorblockA{$^1$MIT,  $^2$CAIDA, $^3$LBNL, $^4$Texas A\&M, $^5$Accolade Technology
}}}
\maketitle

\begin{abstract}
Long range detection is a cornerstone of defense in many operating domains (land, sea, undersea, air, space, ..,).    In the cyber domain, long range detection requires the analysis of significant network traffic from a variety of observatories and outposts.  Construction of anonymized hypersparse traffic matrices on edge network devices can be a key enabler by providing significant data compression in a rapidly analyzable format that protects privacy.  GraphBLAS is ideally suited for both constructing and analyzing anonymized hypersparse traffic matrices.  The performance of GraphBLAS on an Accolade Technologies edge network device is demonstrated on a near worse case traffic scenario using a continuous stream of CAIDA Telescope darknet packets.  The performance for varying numbers of traffic buffers, threads, and processor cores is explored.  Anonymized hypersparse traffic matrices can be constructed at a rate of over 50,000,000 packets per second;  exceeding a typical 400 Gigabit network link.  This performance demonstrates that anonymized hypersparse traffic matrices are readily computable on edge network devices with minimal compute resources and can be a viable data product for such devices.  
\end{abstract}


\section{Introduction}

\begin{figure*}
\center{\includegraphics[width=1.5\columnwidth]{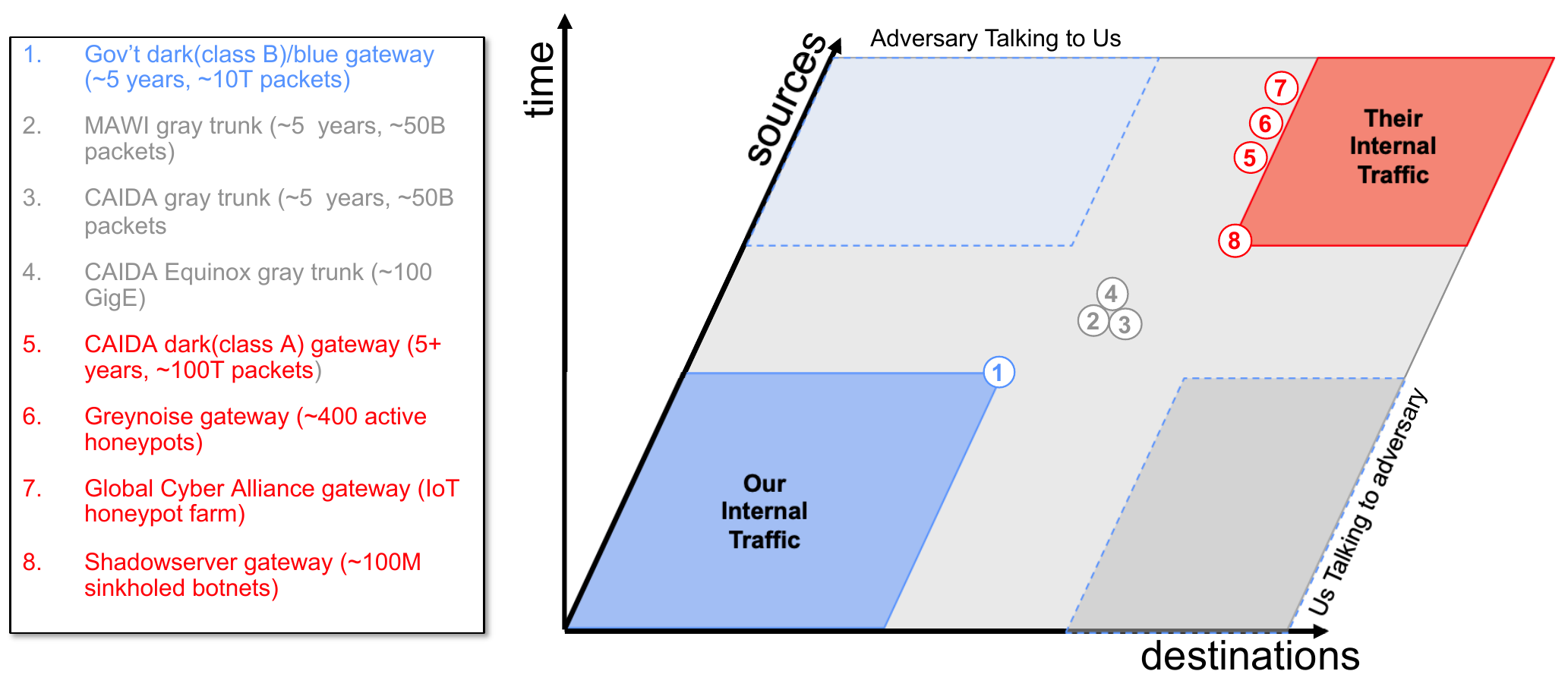}}
      	\caption{{\bf Internet Observatories and Outposts.} Traffic matrix view of the Internet depicting selected observatories and outposts and their notional proximity to various types of network traffic \cite{CAIDA2019, CAIDA2022, GCA2022, Greynoise2022, MAWI2020, Shadowserver2022, kepner2020multi}.}
      	\label{fig:observatories}
\end{figure*}

   The Internet has become as essential as land, sea, air, and space for enabling activities as diverse as commerce, education, health, and entertainment \cite{Cisco2017, Cisco2018-2023}.  Long range detection has been a cornerstone of defense in many operating domains since anitquity \cite{delaney2015perspectives, topouzi2002ancient, shu2021research, cacciotti2010guardian, watson1957three, delaney1990air, geul2017modelling, o2018radar}.  In the cyber domain,   observatories and outposts have been constructed to gather data on  Internet traffic and provide a starting point for exploring long range detection \cite{CAIDA2019, CAIDA2022, GCA2022, Greynoise2022, MAWI2020, Shadowserver2022, kepner2020multi}  (see Figure~\ref{fig:observatories}).  The largest public Internet observatory is the Center for Applied Internet Data Analysis (CAIDA) Telescope that operates a variety of sensors including a continuous stream of packets from an unsolicited darkspace representing approximately 1/256 of the Internet \cite{claffy2000measuring, li2013survey, rabinovich2016measuring, ClaffyClark2020}.  In general, long range detection requires the analysis of significant network traffic from a variety of observatories and outposts \cite{kepner2021zero, weed2022beyond}.
   
     The data volumes, processing requirements, and privacy concerns of analyzing a significant fraction of the Internet have been prohibitive.  The North American Internet generates billions of non-video Internet packets each second \cite{Cisco2017, Cisco2018-2023}.   The GraphBLAS standard  provides significant performance and compression capabilities which improve the feasibility of analyzing these volumes of data \cite{kepner2011graph, kepner2015graphs, kepner16mathematical, buluc17design, kepner2017enabling, yang2018implementing, davis18algorithm, kepner2018mathematics, davis2019algorithm, mattson2019lagraph, cailliau2019redisgraph, davis2019write, aznaveh2020parallel, brock2021introduction, pelletier2021graphblas}.   Specifically, the GraphBLAS is ideally suited for both constructing and analyzing anonymized hypersparse traffic matrices.  Prior work with the GraphBLAS has demonstrated rates of 75 billion packets per second (pps) \cite{kepner202075}, while achieving compressions of 1 bit per packet \cite{kepner2020multi}, and enabling the analysis of the largest publicly available historical archives with over 40 trillion packets \cite{kepner2021spatial}.  Analysis of anonymized hypersparse traffic matrices from a variety of sources has revealed power-law distributions \cite{kepner19hypersparse, kepner2022new}, novel scaling relations \cite{kepner2020multi, kepner2021spatial}, and inspired new models of network traffic \cite{devlin2021hybrid}.
  
GraphBLAS anonymized hypersparse traffic matrices represent one set of design choices for analyzing network traffic.  Specifically, the use case requiring some data on all packets (no down-sampling), high performance, high compression,  matrix-based analysis, anonymization, and open standards.  There are a wide range of alternative graph/network analysis technologies and many good implementations  achieve performance close to the limits of the underlying computing hardware  \cite{tumeo2010efficient, kumar2018ibm, ezick2019combining, gera2020traversing, azad2020evaluation, du2021interactive, acer2021exagraph, blanco2021delayed, ahmed2021online, azad2021combinatorial, koutra2021power}.  Likewise, there are many network analysis tools that focus on providing a rich interface to the full diversity of data found in network traffic \cite{hofstede2014flow, sommer2003bro, lucente2008pmacct}.  Each of these technologies has appropriate use cases in the broad field of Internet traffic analysis.

Sending large volumes of raw Internet traffic to a central location to construct anonymized hypersparse traffic matrices is prohibitive. To meet the goal of providing some data on all packets without down-sampling requires  constructing  the anonymized hypersparse traffic matrices in the network itself in order to realize the full data compression benefits.  The goal of this paper is to explore the viability of this approach by measuring the performance of GraphBLAS on an on edge network device. The performance is measured  on a near worse case traffic scenario using a continuous stream of CAIDA Telescope darknet packets (mostly botnets and scanners) which have an irregular distribution and almost no data payload (i.e., all header).  

The outline of the rest of the paper is as follows.  First, the CAIDA Telescope test data and some basic network quantities  are defined in terms of traffic matrices.   Next, the anonymized hypersparse traffic matrix pipeline is described followed by a description of the experimental setup and implementation.  Finally, the results, conclusions, and directions for further work are presented.

\section{Test Data and Traffic Matrices}

\begin{figure}
\center{\includegraphics[width=0.65\columnwidth]{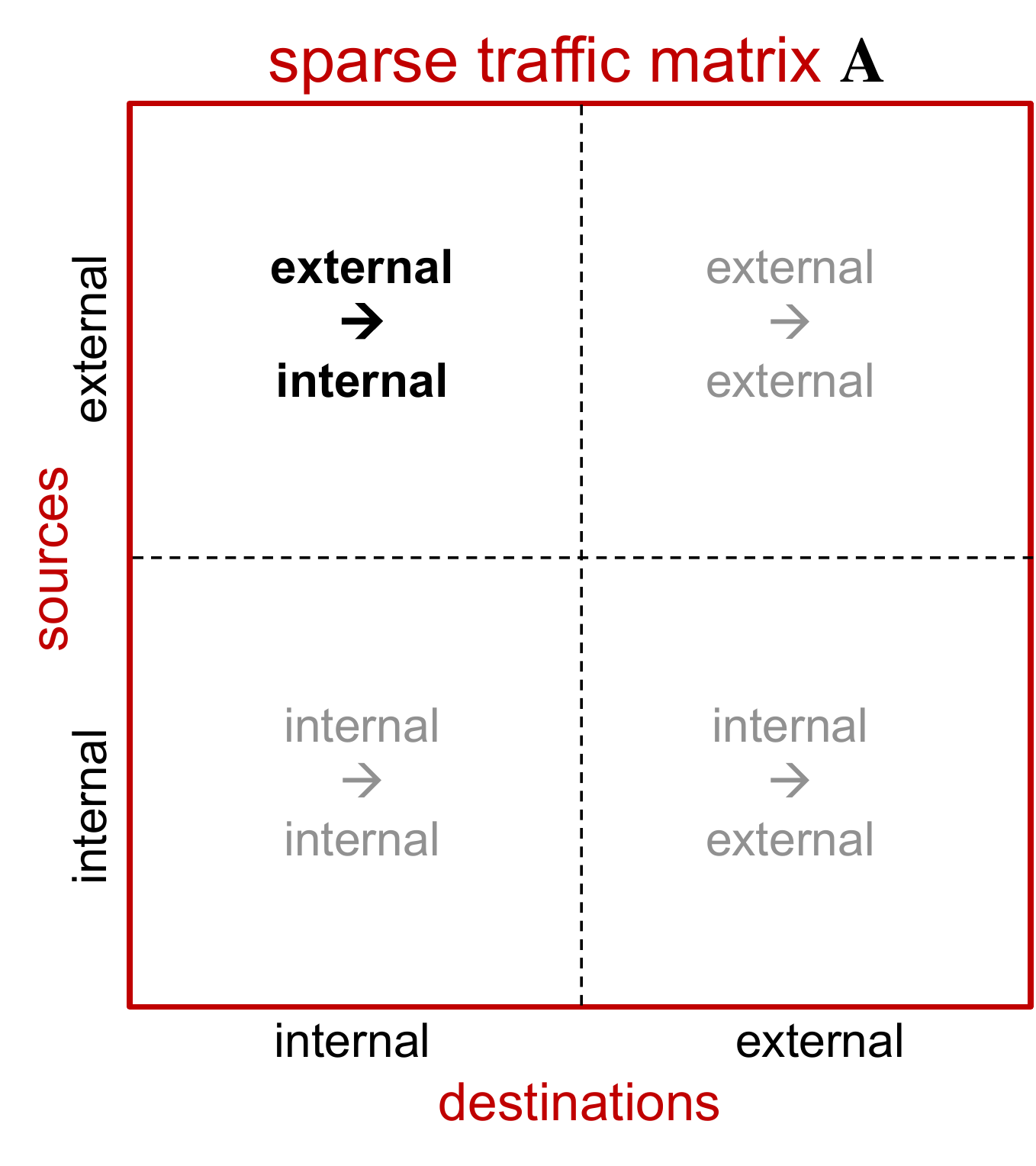}}
      	\caption{{\bf Network traffic matrix.} The traffic matrix can be divided into quadrants separating internal and external traffic.  The CAIDA Telescope monitors a darkspace, so only the upper left (external $\rightarrow$ internal) quadrant will have data.}
      	\label{fig:GatewayTrafficMatrix}
\end{figure}

The test data is drawn from the CAIDA Telescope darknet packets (mostly botnets and scanners) and is a near worse case  with a highly irregular distribution and almost no data payload (i.e., all header).  The CAIDA Telescope monitors the traffic into and out of a set of network addresses providing a natural observation point of network traffic.  These data can be viewed as a traffic matrix where each row is a source and each column is a destination.  The CAIDA Telescope traffic matrix can be partitioned into four quadrants (see Figure~\ref{fig:GatewayTrafficMatrix}).  These quadrants represent different flows between nodes internal and external to the set of monitored addresses.  Because the CAIDA Telescope network addresses are a darkspace, only the upper left (external $\rightarrow$ internal) quadrant will have data.

Internet data  must be handled with care, and CAIDA has pioneered  trusted data sharing best practices that combine anonymizing source and destinations using CryptoPAN \cite{fan2004prefix} with data sharing agreements. These data sharing best practices are the basis of the architecture presented here and include the following principles  \cite{kepner2021zero} 
\begin{itemize}
\item Data is made available in curated repositories
\item Using standard anonymization methods where needed: hashing, sampling, and/or simulation
\item Registration with a repository and demonstration of legitimate research need
\item Recipients legally agree to neither repost a corpus nor deanonymize data
\item Recipients can publish analysis and data examples necessary to review research
\item Recipients agree to cite the repository and provide publications back to the repository
\item Repositories can curate enriched products developed by researchers
\end{itemize}

\begin{figure}
\center{\includegraphics[width=1.0\columnwidth]{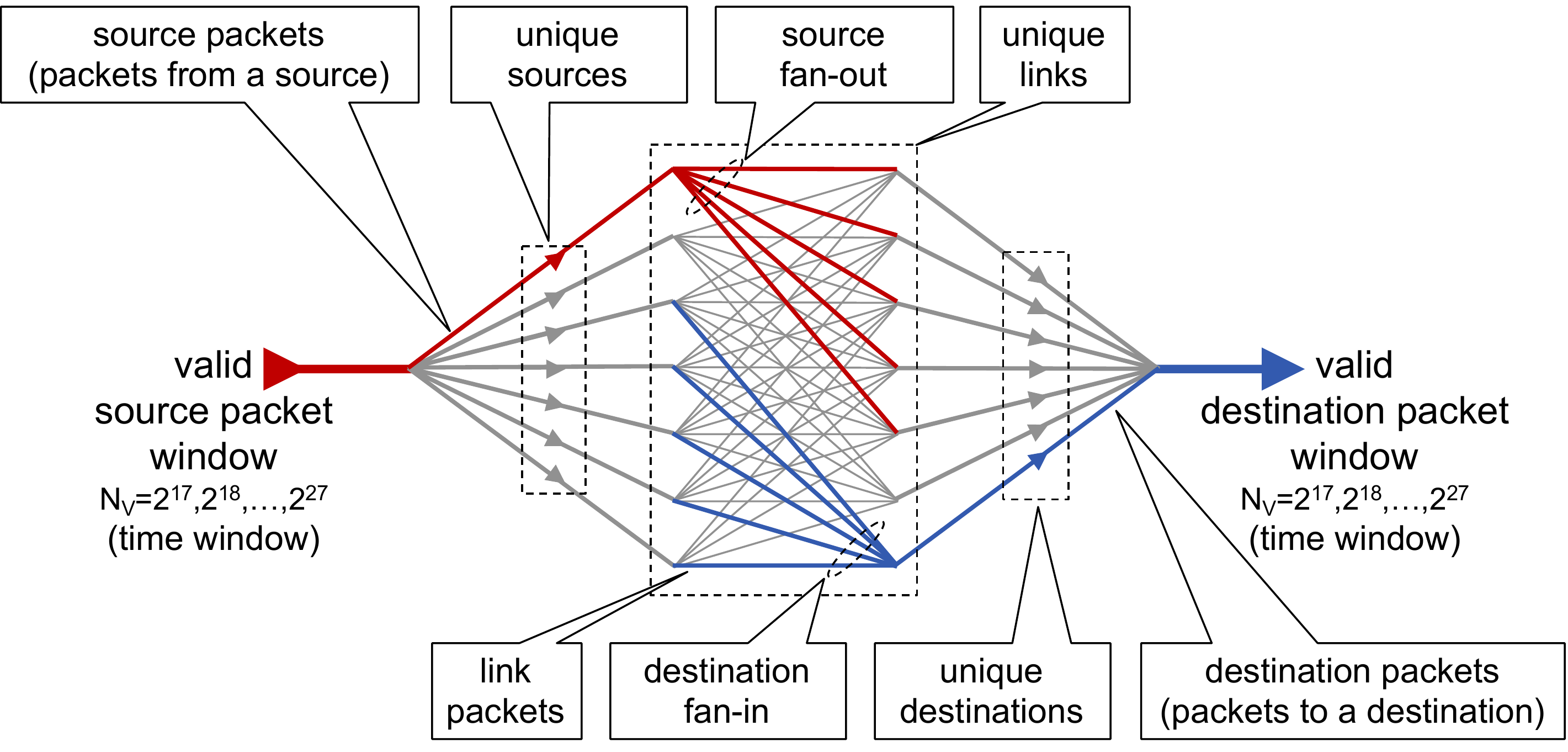}}
      	\caption{{\bf Streaming network traffic quantities.} Internet traffic streams of $N_V$ valid packets are divided into a variety of quantities for analysis: source packets, source fan-out, unique source-destination pair packets (or links), destination fan-in, and destination packets.}
      	\label{fig:NetworkDistribution}
\end{figure}

A primary benefit of constructing anonymized hypersparse traffic matrices with the GraphBLAS is the efficient computation of a wide range of network quantities via matrix mathematics.  Figure~\ref{fig:NetworkDistribution} illustrates essential quantities found in all streaming dynamic networks. These quantities are all computable from anonymized traffic matrices created from the source and destinations found in Internet packet headers.

\begin{table}
\caption{Network Quantities from Traffic Matrices}
\vspace{-0.25cm}
Formulas for computing network quantities from  traffic matrix ${\bf A}_t$ at time $t$ in both summation and matrix notation. ${\bf 1}$ is a column vector of all 1's, $^{\sf T}$  is the transpose operation, and $|~|_0$ is the zero-norm that sets each nonzero value of its argument to 1\cite{karvanen2003measuring}.  These formulas are unaffected by matrix permutations and will work on anonymized data.
\begin{center}
\begin{tabular}{p{1.45in}p{0.9in}p{0.6in}}
\hline
{\bf Aggregate} & {\bf ~~~~Summation} & {\bf ~Matrix} \\
{\bf Property} & {\bf ~~~~~~Notation} & {\bf Notation} \\
\hline
Valid packets $N_V$ & $~~\sum_i ~ \sum_j ~ {\bf A}_t(i,j)$ & $~{\bf 1}^{\sf T} {\bf A}_t {\bf 1}$ \\
Unique links & $~~\sum_i ~ \sum_j |{\bf A}_t(i,j)|_0$  & ${\bf 1}^{\sf T}|{\bf A}_t|_0 {\bf 1}$ \\
Link packets from $i$ to $j$ & $~~~~~~~~~~~~~~{\bf A}_t(i,j)$ & ~~~$~{\bf A}_t$ \\
Max link packets ($d_{\rm max}$) & $~~~~~\max_{ij}{\bf A}_t(i,j)$ & $\max({\bf A}_t)$ \\
\hline
Unique sources & $~\sum_i |\sum_j ~ {\bf A}_t(i,j)|_0$  & ${\bf 1}^{\sf T}|{\bf A}_t {\bf 1}|_0$ \\
Packets from source $i$ & $~~~~~~~\sum_j ~ {\bf A}_t(i,j)$ & ~~$~~{\bf A}_t  {\bf 1}$ \\
Max source packets ($d_{\rm max}$)  & $ \max_i \sum_j ~ {\bf A}_t(i,j)$ & $\max({\bf A}_t {\bf 1})$ \\
Source fan-out from $i$ & $~~~~~~~~~~\sum_j |{\bf A}_t(i,j)|_0$  & ~~~$|{\bf A}_t|_0 {\bf 1}$ \\
Max source fan-out ($d_{\rm max}$) & $ \max_i \sum_j |{\bf A}_t(i,j)|_0$  & $\max(|{\bf A}_t|_0 {\bf 1})$ \\
\hline
Unique destinations & $~\sum_j |\sum_i ~ {\bf A}_t(i,j)|_0$ & $|{\bf 1}^{\sf T} {\bf A}_t|_0 {\bf 1}$ \\
Destination packets to $j$ & $~~~~~~~\sum_i ~ {\bf A}_t(i,j)$ & ${\bf 1}^{\sf T}|{\bf A}_t|_0$ \\
Max destination packets ($d_{\rm max}$) & $ \max_j \sum_i ~ {\bf A}_t(i,j)$ & $\max({\bf 1}^{\sf T}|{\bf A}_t|_0)$ \\
Destination fan-in to $j$ & $~~~~~~~~~~\sum_i |{\bf A}_t(i,j)|_0$ & ${\bf 1}^{\sf T}~{\bf A}_t$ \\
Max destination fan-in ($d_{\rm max}$) & $ \max_j \sum_i |{\bf A}_t(i,j)|_0$ & $\max({\bf 1}^{\sf T}~{\bf A}_t)$ \\
\hline
\end{tabular}
\end{center}
\label{tab:Aggregates}
\end{table}%

The network quantities depicted in Figure~\ref{fig:NetworkDistribution} are computable from anonymized origin-destination traffic  matrices that are widely used to represent network traffic \cite{soule2004identify, zhang2005estimating, mucha2010community, tune2013internet}.  It is common to filter the packets down to a valid set for  any particular analysis.   Such filters may limit particular sources, destinations, protocols, and time windows. To reduce statistical fluctuations, the streaming data should be partitioned so that for any chosen time window all data sets have the same number of valid packets \cite{kepner19streaming}.  At a given time $t$, $N_V$ consecutive valid packets are aggregated from the traffic into a hypersparse matrix ${\bf A}_t$, where ${\bf A}_t(i,j)$ is the number of valid packets between the source $i$ and destination $j$. The sum of all the entries in ${\bf A}_t$ is equal to $N_V$
$$
    \sum_{i,j} {\bf A}_t(i,j) = N_V
$$
Constant packet, variable time samples simplify the statistical analysis of the heavy-tail distributions commonly found in network traffic quantities \cite{kepner19hypersparse, nair2020fundamentals, kepner2022new}.  All the network quantities depicted in Figure~\ref{fig:NetworkDistribution} can be readily computed from ${\bf A}_t$ using the formulas listed in Table~\ref{tab:Aggregates}.  Because matrix operations are generally invariant to permutation (reordering of the rows and columns), these quantities can readily be computed from anonymized data.  Furthermore, the anonymized data can be analyzed by subranges of IPs using simple matrix multiplication.  For a given subrange represented by an anonymized hypersparse diagonal matrix ${\bf A}_r$, where ${\bf A}_r(i,i) = 1$ implies  source/destination $i$ is in the range, the traffic within the subrange can be computed via: ${\bf A}_r {\bf A}_t  {\bf A}_r$. Likewise, for additional privacy guarantees that can be implemented at the  edge, the same method can be used to exclude a range of data from the traffic matrix
$$
     {\bf A}_t - {\bf A}_r {\bf A}_t  {\bf A}_r
$$

The contiguous nature of these data allows the exploration of a wide range of packet windows from $N_V = 2^{17}$ (sub-second) to $N_V = 2^{27}$ (minutes), providing a unique view into how network quantities depend upon time.  These observations provide new insights into normal network background traffic that could be used for anomaly detection, AI feature engineering, polystore index learning, and testing theoretical models of streaming networks \cite{elmore2015demonstration, kraska18case, do20classifying}.

\begin{figure}
\center{\includegraphics[width=1.0\columnwidth]{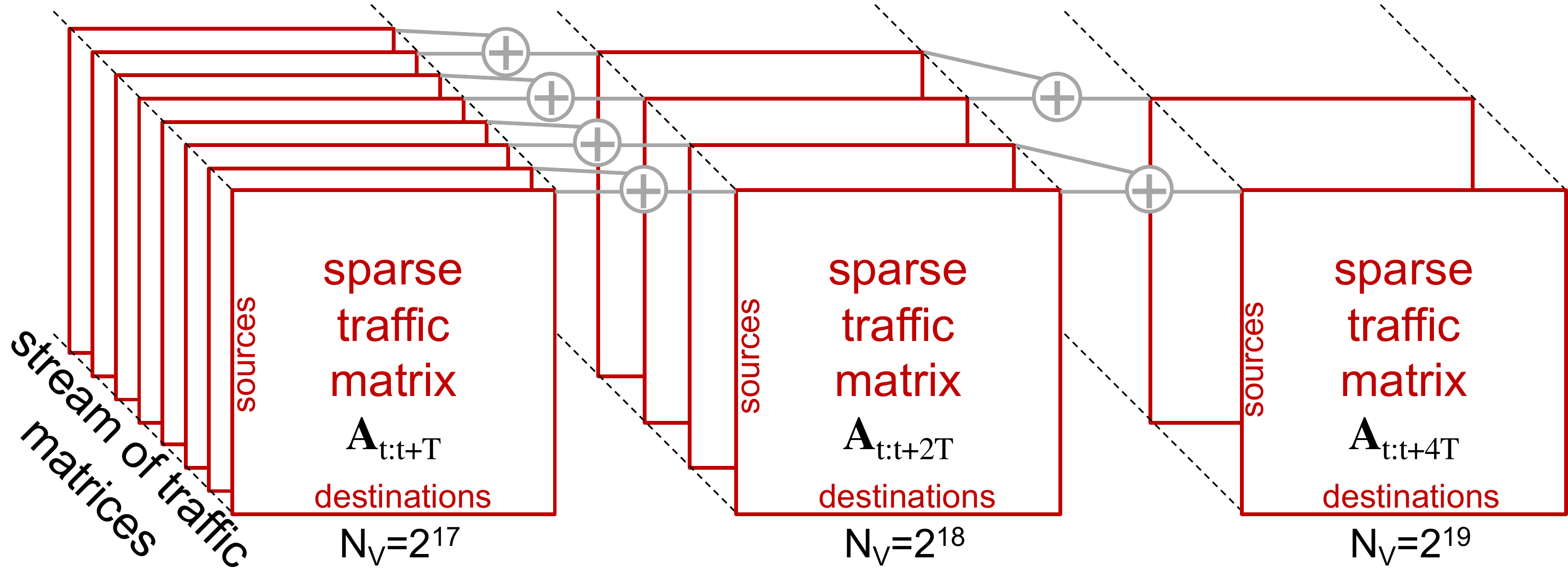}}
      	\caption{{\bf Multi-temporal streaming traffic matrices.} Efficient computation of network quantities on multiple time scales can be achieved by hierarchically aggregating data in different time windows.}
      	\label{fig:MultiTemporalMatrix}
\end{figure}

Network traffic is dynamic and exhibits varying behavior on a wide range of time scales.  A given packet window size $N_V$ will be sensitive to phenomena on its corresponding timescale.  Determining how network quantities scale with $N_V$ provides insight into the temporal behavior of network traffic.   Efficient computation of network quantities on multiple time scales can be achieved by hierarchically aggregating data in different time windows \cite{kepner19streaming}.  Figure~\ref{fig:MultiTemporalMatrix} illustrates a binary aggregation of  different streaming traffic matrices.   Computing each quantity at each hierarchy level eliminates redundant computations that would be performed if each packet window was computed separately.  Hierarchy also ensures that most computations are performed on smaller matrices residing in faster memory.  Correlations among the matrices mean  that adding two matrices each with $N_V$ entries results in a matrix with fewer than $2N_V$ entries, reducing the relative number of operations as the matrices grow.

One of the important capabilities of the SuiteSparse GraphBLAS library is direct support of hypersparse matrices where the number of nonzero entries is significantly less than either dimensions of the matrix.  If the packet source and destination identifiers are drawn from a large numeric range, such as those used in the Internet protocol, then a hypersparse representation of ${\bf A}_t$ eliminates the need to keep track of additional indices and can significantly accelerate the computations \cite{kepner202075}.

\section{Hypersparse Matrix Pipeline}

\begin{figure*}
\center{\includegraphics[width=1.5\columnwidth]{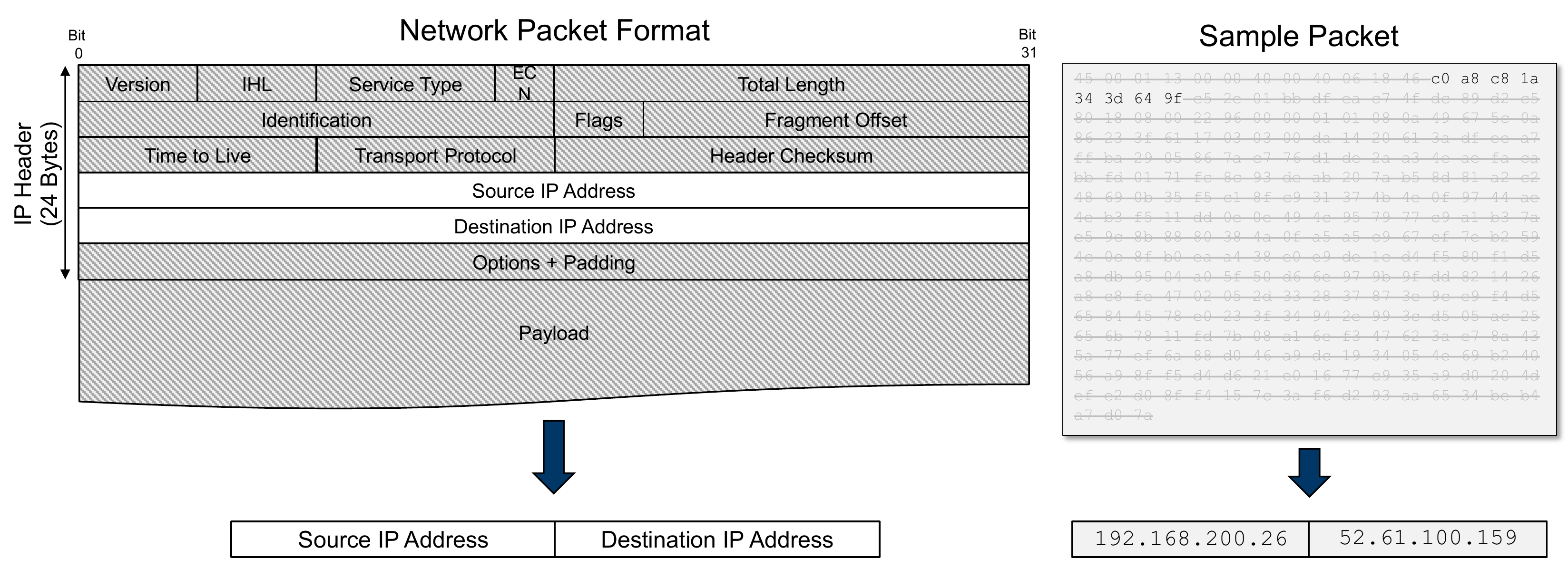}}
      	\caption{{\bf Network Packet Description.} A network packet consists of a header and a payload.  To avoid downsampling and minimize privacy concerns only the source and destination are selected.}
      	\label{fig:packet}
\end{figure*}

\begin{figure*}
\center{\includegraphics[width=1.5\columnwidth]{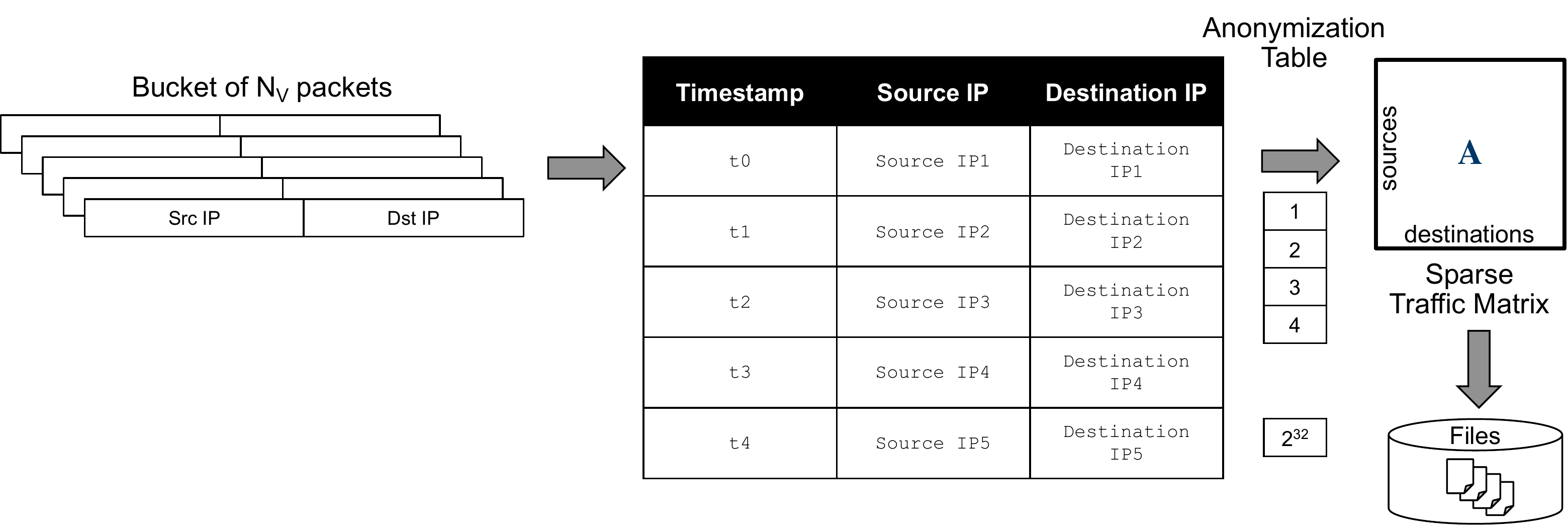}}
      	\caption{{\bf Anonymized Hypersparse Matrix Pipeline.} Continuous sequences of $N_V=2^{17}$ packets are extracted from packet headers, anonymized, formed into a GraphBLAS hypersparse matrix, serialized, and saved in groups of 64 GraphBLAS matrices to a UNIX TAR file.}
      	\label{fig:pipeline}
\end{figure*}

The aforementioned analysis goals set the requirements for the GraphBLAS hypersparse traffic matrix pipeline.  Specifically, the  compression benefits are maximized if the GraphBLAS hypersparse traffic matrix is constructed in the network as close to the network traffic as possible as this minimizes the amount of  data that needs to be sent over the network.  In addition, collection at the network source allows the data owner to construct and own the anonymization scheme and only share anonymized data under trusted data sharing agreements with the parties tasked with analyzing the data \cite{pisharody2021realizing}.

The first step in the GraphBLAS hypersparse traffic matrix pipeline is to capture a packet, discard the data payload, and extract the source and destination Internet Protocol (IP) addresses (Figure~\ref{fig:packet}).   For the purposes of the current performance testing, only IPv4 packets are used which are stored as 32 bit unsigned integers.  Collections of $N_V = 2^{17}$ consecutive packets are then each anonymized using a cryptoPAN generated anonymization table.  The resulting anonymized source and destination IPs are then used to construct a $2^{32}{\times}2^{32}$ hypersparse GraphBLAS matrix.  64 consecutive hypersparse GraphBLAS matrices are each serialized in compressed sparse rows (CSR) format with LZ4 compression and saved to a UNIX TAR file (Figure~\ref{fig:pipeline}).  The TAR files can be further compressed using other compressing methods (if desired) and then transmitted to the appropriate parties tasked for analysis.   For example, standard gzip compression reduces file size by 40\% but also reduces performance by 80\%.

\section{Implementation}

\begin{figure}
\center{\includegraphics[width=1.0\columnwidth]{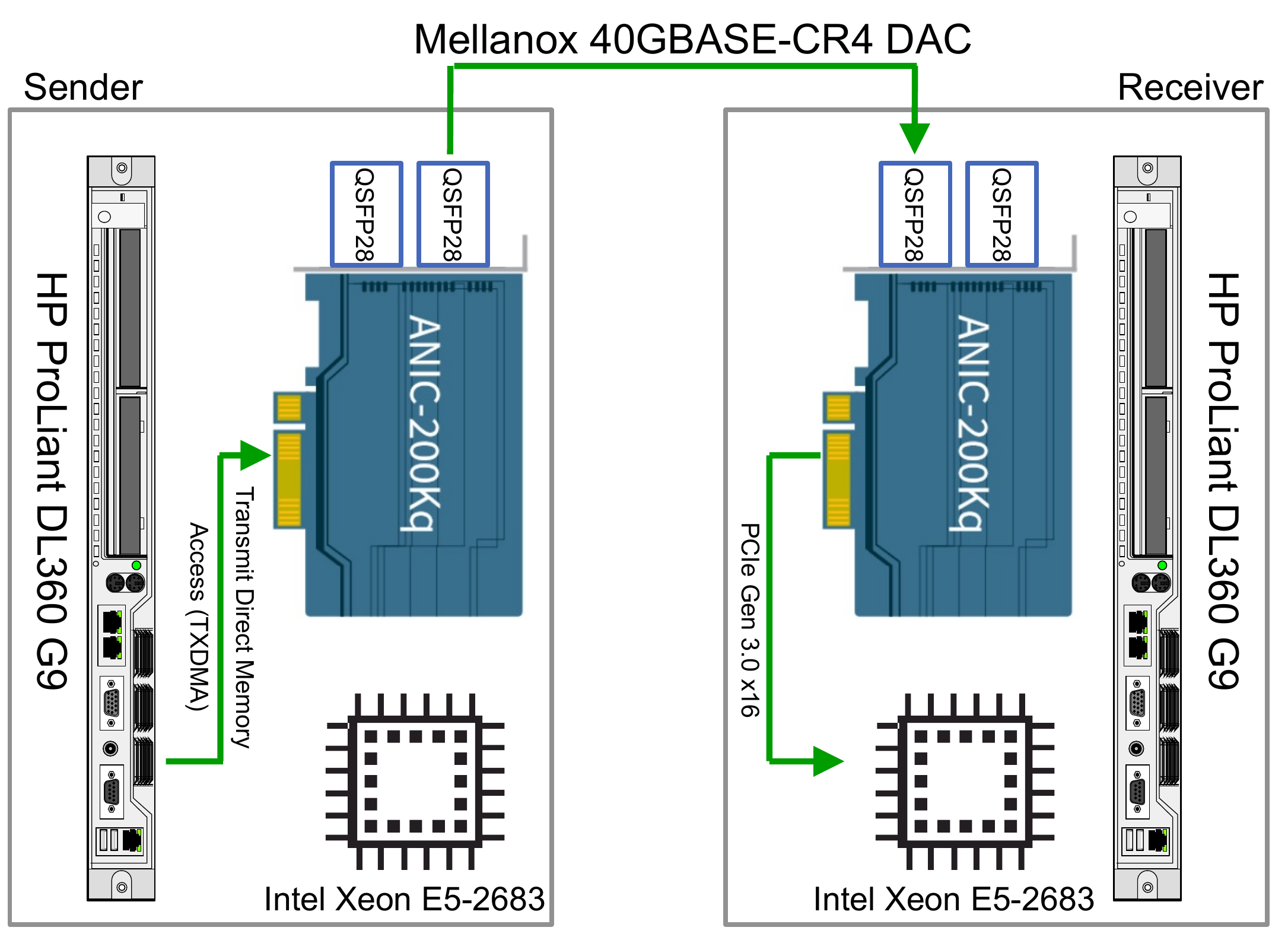}}
      	\caption{{\bf Experimental Setup.} Test system consisted of two compute nodes each with an Accolade card connected over a network. The sender node reads CAIDA Telescope packet data into memory and then the Accolode card sends the data directly from memory over the network to the receiver node. The Accolade card on the receiver takes the data off the network, places the data in a hardware ring buffer, and makes the data available to be processed by the receiver processor.}
      	\label{fig:setup}
\end{figure}

Effective implementation of the GraphBLAS pipeline requires that the anonymization, creation, and saving of the resulting files can keep up with the data rates of typical high bandwidth links.  To measure this performance two Accolade Technology ANIC-200Kq dual port 100 gigabit flow classification and shunting adapters were installed into the PCIe slots of two HP Proliant DL360 G9 servers (Figure~\ref{fig:setup}).   These servers were connected via a Mellanox 40 gigabit network connection.  ANIC-200Kq cards are capable of a wide range of analysis techniques, this experiment only used their data transmission, shunting, and buffering capabilities.

The C implementation of the GraphBLAS hypersparse traffic matrix pipeline shown in Figures~\ref{fig:packet} and \ref{fig:pipeline} was run on dual Intel Xeon E5-2683 processors in the receiver server.  The Accolade card on the receiver server collects packets in ring buffers, the number of which can be set at initialization.  Using C pthreads \cite{nichols1996pthreads}  an Accolade worker thread is assigned to each Accolade hardware ring.  Within each Accolade worker thread, a block of $2^{23}$ IPv4 packets are collected and a GraphBLAS worker thread is spawned to process the block in subblocks of $2^{17}$ packets. Each sublock is anonymized  using  a cryptoPAN generated anonymization vector and the resulting anonymized sources and destinations are used to construct a GraphBLAS matrix.  The matrix is the serialized and appended to a TAR buffer, which is saved to a file after all 64 subblocks have been processed.  A more detailed outline of the code is as follows:

\noindent \_\_\_\_\_\_\_\_\_\_\_\_\_\_\_\_\_\_\_\_\_\_\_\_\_\_\_\_\_\_\_\_\_\_\_\_\_\_\_\_\_\_\_\_\_\_\_\_\_\_
{\small
\noindent \underline{Main Thread}
\begin{itemize}
\item Set number of Accolade hardware rings
\item Load $2^{32}$ entry IPv4 anonymization table
\item For each Accolade hardware ring
\begin{itemize}
\item Launch \underline{Accolade Worker} thread
\end{itemize}
\end{itemize}
\noindent \underline{Accolade Worker}
\begin{itemize}
\item Create libpcap handle to Accolade device
\item Allocate 64MB buffer for packet processing
\item For each packet
\begin{itemize}
\item Retrieve packet from Accolade device buffers
\item Append source and destination IP addresses to buffer
\item If buffer has $2^{23}$ packets
\begin{itemize}
\item[$\circ$] Launch \underline{GraphBLAS Worker} thread with pointer to packet buffer
\item[$\circ$] Allocate new 64MB buffer for packet processing
\end{itemize}
\end{itemize}
\end{itemize}
\noindent \underline{GraphBLAS Worker}
\begin{itemize}
\item Initialize GraphBLAS
\item For each subblock of $2^{17}$ packets
\begin{itemize}
\item Create new GraphBLAS matrix
\item Create row, column and value vectors
\item For each packet in subblock
\begin{itemize}
\item [$\circ$] Lookup in anonymization table
\item [$\circ$] Insert into row, column and value vectors
\end{itemize}
\item Build GraphBLAS matrix from row, column and value vectors; summing duplicate entries
\item Serialize and compress GraphBLAS matrix
\item Append to TAR buffer
\end{itemize}
\item Write TAR buffer to file
\end{itemize}
}
\vspace{-0.4cm}
\noindent \_\_\_\_\_\_\_\_\_\_\_\_\_\_\_\_\_\_\_\_\_\_\_\_\_\_\_\_\_\_\_\_\_\_\_\_\_\_\_\_\_\_\_\_\_\_\_\_\_\_

\section{Results}

\begin{figure}
\center{\includegraphics[width=1.0\columnwidth]{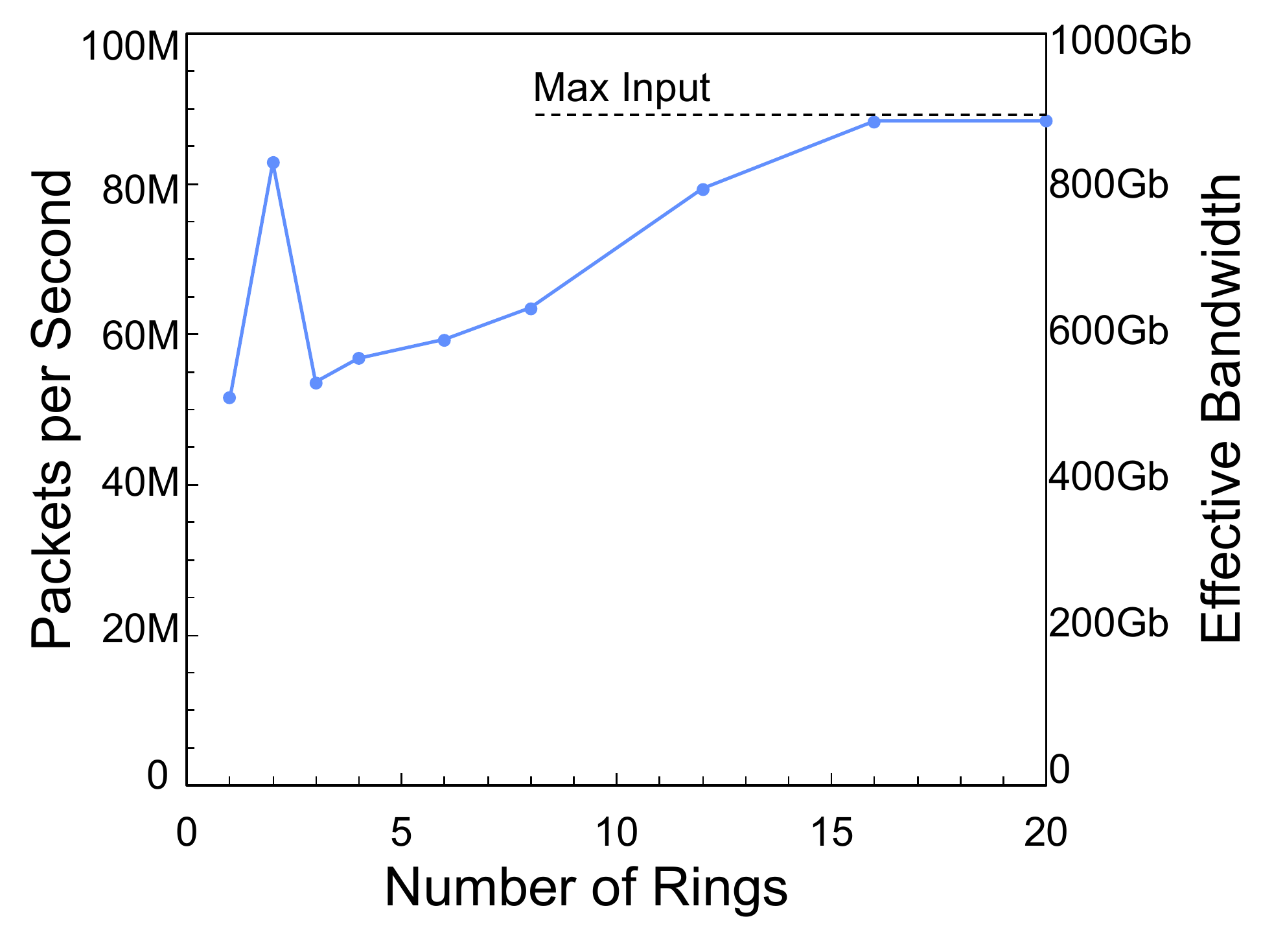}}
      	\caption{{\bf Performance Results.}  Packets per second processed versus the number of hardware rings used. The packet performance can be converted into an estimated equivalent bandwidth using a representative packets size of 10,000 bits per packet (see right vertical axis)}
      	\label{fig:performance}
\end{figure}

\begin{figure}
\center{\includegraphics[width=1.0\columnwidth]{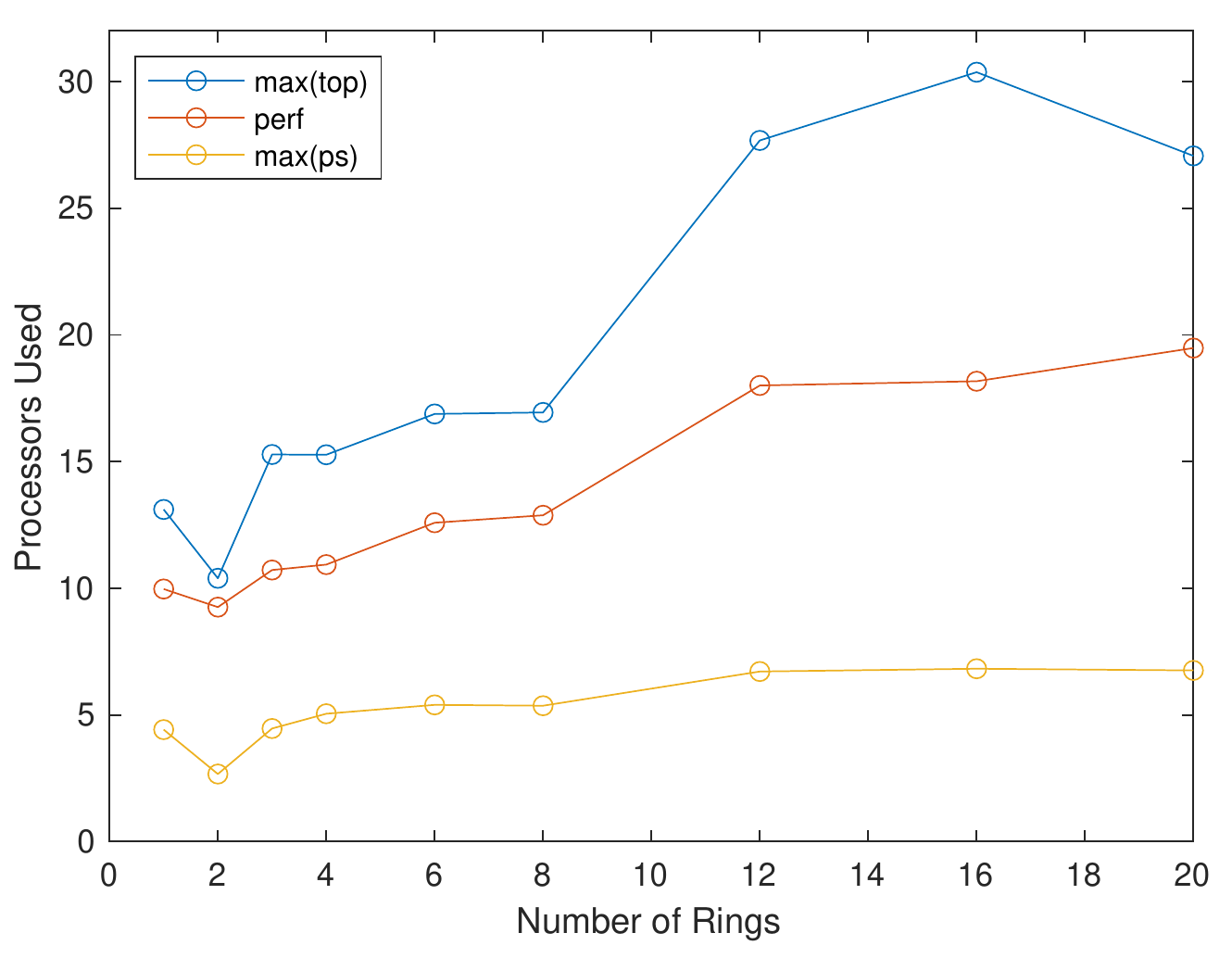}}
      	\caption{{\bf Processors Used.}  Processors used versus the number of hardware rings as determined by the max load of the Linux {\sf top} command, average load of the Linux {\sf perf} command, and the max load of the Linux {\sf ps} command.}
      	\label{fig:proc}
\end{figure}

Using the experimental setup shown in Figure~\ref{fig:setup}, a number of performance experiments were conducted.  In these experiments the sender server would load $100{\times}2^{23}$ CAIDA Telescope packets into the sender Accolade card and sends the packets at defined rate over the network to the receiver server where the receiver Accolade card  loads the packets into its hardware rings.  Likewise, on the receiver server the GraphBLAS hypersparse traffic matrix pipeline described in the previous section was executed.  The cryptoPAN anonymization table was created offline, stored, and loaded at startup.  The reason for this is that the single core cryptoPAN performance is approximately 700,000 IP address per second.  The $2^{32}$ entry cryptoPAN anonymization lookup table dramatically speeds up the performance.  Generation of the table is readily run in parallel and can be generated in a few seconds (if desired). The rate of packets being sent was adjusted to achieve the maximum rate without dropped packets, which indicated that the GraphBLAS hypersparse traffic matrix pipeline was able to keep up with the incoming traffic.  The number of hardware rings were varied which increase the number of threads/cores being used.  

CAIDA Telescope data are a near worse case scenario because  almost all of the packets have no payload and very few packets use the same source and destination connection so the resulting hypersparse traffic matrices have very few entries that are greater than 1.  An advantage of the few payloads  in the CAIDA Telescope data is that it allows the emulation of packet streams that are representative of much higher bandwidth networks than the current experimental setup is capable of.

Figure~\ref{fig:performance} shows the performance in terms packets per second versus the number of hardware rings used.  GraphBLAS matrix construction is highly cache sensitive due to the underlying index sorting required.  The  performance of a single ring using a few processing threads/cores is over 50M packets per second.  The performance increases significantly using two rings, and then drops with 3 rings because of cache effects.  Using 16 rings makes up for the cache effects with increased parallelism and the performance reaches the maximum number of packets the sender server can send to the receiver (approximately 88M packets per second).  The packet performance can be converted into an estimated equivalent bandwidth using a representative packet size of 10,000 bits per packet (see Figure~\ref{fig:performance} right vertical axis).  The performance measurements indicate that a standard server is a capable of constructing anonymized hypersparse traffic matrices at a rate above that corresponding to a typical 400 Gigabit network link.

Figure~\ref{fig:proc} shows the processors used as computed by the max load of the Linux {\sf top} command, average load of the Linux {\sf perf} command, and the max load of the Linux {\sf ps} command.  {\sf top} is a point in time measurement of utilization at the precise moment of polling and was sampled at 1 second intervals.  {\sf perf} is an estimated average of the processor load throughout the process lifetime.   {\sf ps} reports percentage of time spent running during the entire lifetime of the process and was sampled at 1 second intervals.  Combined these results indicate that the peak performance achieved using 2 rings (see Figure~\ref{fig:performance}) required only a handful of cores.

\section{Conclusions and Future Work}

For many operating domains (land, sea, undersea, air, space, ..,) long range detection is a cornerstone of defense.  Long range detection in the cyber domain requires significant network traffic to be analyzed from a variety of observatories and outposts.  Construction of anonymized hypersparse traffic matrices on edge network devices can be a key enabler by providing significant data compression in a rapidly analyzable format that protects privacy.  Constructing and analyzing anonymized hypersparse traffic matrices are operations ideally suited to the GraphBLAS high performance library.    Using an Accolade Technologies edge network device the performance of the GraphBLAS is demonstrated on a near worse case traffic scenario using a continuous stream of CAIDA Telescope darknet packets.  The performance was explored by varying  the number of traffic ring buffers, which are proportional to the number of threads and processor cores used.   Rates of over 50,000,000 packets per second for constructing anonymized hypersparse traffic matrices were achieved which exceeds  a typical 400 Gigabit network link.

This performance demonstrates that anonymized hypersparse traffic matrices are readily computable on edge network devices with minimal compute resources and can be a viable data product for such devices.  This work suggests a variety of future directions that could be pursued (1) exploring additional network cards; (2) develpoing the appropriate key management architecture for multiple observatories and outposts; (3) analysis of spatial temporal patterns in anonymized traffic matrices to identify adversarial activities; (4) cross-correlation of data from different observatories and outposts; (5) development of AI algorithms for classification of background traffic; (6) creation of underlying models of traffic.

\section*{Acknowledgments}

The authors wish to acknowledge the following individuals for their contributions and support: Bob Bond, Stephen Buckley, Ronisha Carter, Cary Conrad, Alan Edelman, Tucker Hamilton, Jeff Gottschalk, Nathan Frey, Chris Hill, Mike Kanaan, Tim Kraska, Andrew Morris, Charles Leiserson, Dave Martinez, Mimi McClure, Joseph McDonald, Sandy Pentland, Christian Prothmann, John Radovan, Steve Rejto, Daniela Rus, Matthew Weiss, Marc Zissman.

\bibliographystyle{ieeetr}
\bibliography{GraphBLASonEdge}

\end{document}